\documentclass[apjl]{emulateapj}

\usepackage{natbib}
\usepackage{amsmath}
\usepackage{amssymb}
\usepackage[bf, sf, FIGTOPCAP, nooneline, tight]{subfigure}
\usepackage{graphicx}
\usepackage{dcolumn}
\usepackage{bm}
\usepackage[usenames,dvipsnames]{color}
\usepackage[colorlinks=true, linkcolor=BrickRed, citecolor=Blue, urlcolor=Blue, filecolor=Blue]{hyperref}
\usepackage{booktabs}
\usepackage{hhline}
\usepackage{epsf}
\usepackage{color}
\usepackage{epsf}
\usepackage{psfrag}
\usepackage{epsfig}
\usepackage{epstopdf}

\usepackage{multirow}
\usepackage{bm}
\usepackage{latexsym}
\usepackage{ifsym}
\usepackage{tabularx}

\usepackage{threeparttable}

\newcommand{\be}{\begin{equation}}
\newcommand{\ee}{\end{equation}}
\def\beq{\begin{equation}}
\def\eeq{\end{equation}}
\def\ber{\begin{eqnarray}}
\def\eer{\end{eqnarray}}

\def \lleq {\lower0.9ex\hbox{ $\buildrel < \over \sim$} ~}
\def \ggeq {\lower0.9ex\hbox{ $\buildrel > \over \sim$} ~}

\def\deg{\ifmmode^\circ\else$^\circ$\fi}

\newcommand{\rma}{\rho_m}
\newcommand{\rr}{\rho_r}
\newcommand{\rL}{\rho_\Lambda}
\newcommand{\dG}{\dot{G}}

\newcommand{\drL}{\dot{\rho}_\Lambda}
\newcommand{\dH}{\dot{H}}
\newcommand{\CC}{\Lambda}

\newcommand{\Omo}{\Omega_{m}}

\newcommand{\ORo}{\Omega_{r}}

\newcommand{\OLo}{\Omega_{\Lambda}}

\newcommand{\rmo}{\rho_{m 0}}

\newcommand{\rmr}{\rho_m}

\newcommand{\rR}{\rho_r}

\newcommand{\rLo}{\rho_{\CC 0}}

\newcommand{\nueff}{\nu_{\rm eff}}
\newcommand{\nueffp}{\nu_{\rm eff}'}

\newcommand{\rRo}{\rho_{r 0}}

\begin{document}

\title{First evidence of running cosmic vacuum: challenging the concordance model}

\author{Joan Sol\`{a}\altaffilmark{1}, Adri\`{a} G\'omez-Valent\altaffilmark{1}, Javier de Cruz P\'erez\altaffilmark{1}
}

\altaffiltext{1}
{Departament de F\'isica Qu\`antica i Astrof\'isica, and Institute of Cosmos Sciences, Univ. de Barcelona, Av. Diagonal 647, E-08028 Barcelona, Catalonia, Spain}


\email{sola@fqa.ub.edu}
\email{adriagova@fqa.ub.edu}
\email{decruz@fqa.ub.edu}

\begin{abstract}
Despite the fact that a rigid $\CC$-term is a fundamental building block of the concordance $\CC$CDM model, we show that a large class of cosmological scenarios with dynamical vacuum energy density $\rL$ and/or gravitational coupling $G$, together with a possible non-conservation of matter, are capable of seriously challenging the traditional phenomenological success of the $\CC$CDM. In this paper, we discuss these ``running vacuum models'' (RVM's), in which $\rL=\rL(H)$  consists of a nonvanishing constant term and a series of powers of the Hubble rate. Such generic structure is potentially linked to the quantum field theoretical description of the expanding Universe. By performing an overall fit to the cosmological observables SNIa+BAO+$H(z)$+LSS+BBN+CMB (in which the WMAP9, Planck 2013 and Planck 2015 data are taken into account), we find that the class of RVM's appears significantly more favored than the $\CC$CDM, namely at an unprecedented level of $\gtrsim4.2\sigma$. Furthermore, the Akaike and Bayesian information criteria confirm that the dynamical RVM's are strongly preferred as compared to the conventional rigid $\CC$-picture of the cosmic evolution.

\end{abstract}

\keywords{dark energy---dark matter---large-scale structure of universe}

\section{Introduction}
As of about twenty years ago dark energy (DE) has become  an observational fact of the first magnitude in physics\,\citep{SNIaRiess,SNIaPerl}  and the most recent observations do not cease to corroborate its existence as the prime cause for the acceleration of the Universe\,\citep{Planck2015,PlanckDE2015}. Next year it will be the centenary of the cosmological constant (CC) term, $\CC$, in Einstein's equations. The $\CC$-term is usually considered the simplest possible explanation for the DE and is an essential ingredient of the so-called concordance or $\Lambda$CDM model. On the theoretical side, however, the explanation is not so easy. In quantum field theory (QFT) the large value predicted for $\CC$, or equivalently for the corresponding vacuum energy density associated to it, $\rL=\CC/(8\pi G)$ ($G$ being Newton's gravitational coupling), as compared to the measured one generates the old CC problem, for reviews see \,\citep{Weinberg,CCP1,CCP2,CCP3, JSPRev2013}. It is probably one of the most fundamental and unsolved conundrums of theoretical physics. The acclaimed finding of the Higgs boson of the standard model of particle physics at the LHC actually bolsters even more the problem since it adopts a more experimental basis. In fact, the associated electroweak (EW) vacuum energy density reads $|\rho_{\rm EW}|\sim M_H^2/G_F$, where $M_H\simeq 125$ GeV is the measured Higgs boson mass and $G_F$ is Fermi's constant. Thus, in this most realistic situation, the CC problem appears on comparing the two quantitites $\rho_{\rm EW}\sim 10^9$ GeV$^4$ to $\rL\sim 10^{-47}$GeV$^4$, which differ by an appalling amount of 56 orders of magnitude.

With such a state of affairs, cosmologists have  felt motivated to look for many other sources of DE beyond $\CC$. For example, scalar fields in cosmology, $\phi$, have been used since long ago, most conspicuously in the context of Brans-Dicke theories\,\citep{BD61}, where $G\propto 1/\phi(t)$, and subsequently in general scalar-tensor theories. But soon also played a role as a strategy to endow the vacuum and the CC of some time dependence in a QFT context, $\CC=\CC(\phi(t))$, and in some cases with the purpose to adjust dynamically its value. Some of the old approaches to the CC problem from the scalar field perspective are the works by \cite{EndoFukui197782,Fujii82,Dolgov83,Abbott85,Zee85,Barr87,Ford87,Weiss87}.  Among the proposed dynamical mechanisms, let us mention the cosmon model\,\citep{PSW}, which was subsequently discussed in detail  by\,\cite{Weinberg}. In all cases, a more or less obvious form of fine tuning underlies the adjusting mechanisms. For this reason scalar fields were later used mostly to ascribe a possible evolution to the vacuum energy with the hope to explain the cosmic coincidence problem, giving rise to the notion of quintessence and the like, cf. \citep{PeeblesRatra88a,PeeblesRatra88b,Wetterich88,Wetterich95,Caldwell98,ZlatevWangSteinhardt99,Amendola2000}, among many other alternatives.  See e.g. the reviews by\,\cite{CCP2,CCP3,Copeland} and the book by\,\cite{DEBook}, and references therein. Let us also mention some of the old cosmological models based on attributing a phenomenological direct time-dependence to the the CC term, $\CC=\CC(t)$, without an obvious relation to scalar fields, see e.g. \,\cite{OzerTaha8687,Bertolami86,Freese87,Carvalho92,Waga93,LimaMaia93,ArcuriWaga94,Arbab97}. Many other works are available in the literature,
the reader can consult the reviews e.g. by \cite{Overduin98}, \cite{Vishwakarma01}, \cite{JSPRev2013}, and references therein.

The old CC problem is a problem of fundamental nature that shows the profound interconnection among different branches of modern physics. Some of the above old works aimed at solving the problem at a time when it was thought that $\CC=0$, so it was expected that some symmetry or some dynamical mechanism could help. But the task  became much harder when it was realized that the CC value is nonvanishing and actually very small in particle physics units ($\rL\sim 10^{-47}$ GeV$^4$).

In this work we will not face the CC problem as such, not even the cosmic coincidence problem. Our main aim is much more modest. Taking into account the current amount and quality of the cosmological data on SNIa+BAO+$H(z)$+LSS+BBN+CMB,  we wish to put to the test the possibility that the $\CC$-term and its associated vacuum energy density, $\rL=\CC/(8\pi G)$, could actually be dynamical (``running'') quantities whose rhythms of variation might be linked to the Universe's expansion rate, $H$. The idea is to check if this possibility helps to improve the description of the overall cosmological data as compared to the rigid assumption $\CC=$const. inherent to the concordance $\CC$CDM model. For the class of models being considered we do not make any direct association of the $\CC$ and $G$  running with the dynamical evolution of scalar fields.  The proposal being investigated here can be motivated in QFT in curved spacetime (cf.\,\,\cite{JSPRev2013,SolGom2015}  and references therein) and we want to show that it can be currently tested. Although a simple Lagrangian description of these models at the level of standard scalar fields is not available, attempts have been made in the literature\,\citep{Fossil07,JSPRev2013} and in any case this is of course something that one would eventually hope to find. There is, however, no guarantee that such description is possible in terms of a simple local action\,\citep{Fossil07}.

Our main aim here is phenomenological. We will argue upon carefully confronting theory and observations that the idea of running vacuum models (RVM's) can be highly competitive, if not superior, to the traditional $\CC$CDM framework.
The first serious indications of dynamical vacuum energy (at the $\sim 3\sigma$ c.l.) were reported in \cite{ApJ1}. Earlier comprehensive studies hinted also at this possibility but remained at a lower level of significance, see e.g. \cite{BPS2009,Grande2011,GomSolBas2015}\footnote{ Recent claims that the $\Lambda$CDM may not be the best description of our Universe can also be found in e.g. \cite{SahniShafielooStarobinsky}, \cite{Ding} and \cite{Zheng}; see, however, Section 3, point S4).}. Remarkably, in the present work the reported level of evidence is significantly higher than in any previous work in the literature (to the best of our knowledge).
While Occam's razor says that ``Among equally competing models describing the same observations, choose the simplest one'', the point we wish to stress here is that the RVM's are able to describe the current observations better than the $\CC$CDM, not just alike.  For this reason we wish to make a case for the RVM's, in the hope that they could shed also some new light on the CC problem, e.g. by motivating further theoretical studies on these models or related ones.

The plan of the paper is as follows. In section 2 we describe the different types of running vacuum models (RVM's) that will be considered in this study. In section 3 we fit these models to a large set of cosmological data on distant type Ia supernovae (SNIa), baryonic acoustic oscillations (BAO's), the known values of the Hubble parameter at different redshift points, the large scale structure (LSS) formation data, the BBN bound on the Hubble rate, and, finally, the CMB distance priors from WMAP and Planck. We include also a fit of the data with the standard XCDM parametrization, which serves as a baseline for comparison. In section 4 we present a detailed discussion of our results, and finally in section 5 we deliver our conclusions.


\begin{table*}
 \caption{Best-fit values for  $\CC$CDM, XCDM and the various running vacuum models (RVM's) using the Planck 2015 results and the full data set S1-S7}
\begin{center}
\resizebox{1\textwidth}{!}{
\begin{tabular}{| c | c |c | c | c | c | c | c | c | c |}
\multicolumn{1}{c}{Model} &  \multicolumn{1}{c}{$h$} &  \multicolumn{1}{c}{$\omega_b= \Omega_b h^2$} & \multicolumn{1}{c}{{\small$n_s$}}  &  \multicolumn{1}{c}{$\Omega_m$}&  \multicolumn{1}{c}{{\small$\nu_{eff}$}}  & \multicolumn{1}{c}{$\omega$}  &
\multicolumn{1}{c}{$\chi^2_{\rm min}/dof$} & \multicolumn{1}{c}{$\Delta{\rm AIC}$} & \multicolumn{1}{c}{$\Delta{\rm BIC}$}\vspace{0.5mm}
\\\hline
$\Lambda$CDM  & $0.693\pm 0.003$ & $0.02255\pm 0.00013$ &$0.976\pm 0.003$& $0.294\pm 0.004$ & - & $-1$  & 90.44/85 & - & - \\
\hline
XCDM  & $0.670\pm 0.007$& $0.02264\pm0.00014 $&$0.977\pm0.004$& $0.312\pm0.007$ & - &$-0.916\pm0.021$  & 74.91/84 & 13.23 & 11.03 \\
\hline
A1  & $0.670\pm 0.006$& $0.02237\pm0.00014 $&$0.967\pm0.004$& $0.302\pm0.005$ &$0.00110\pm 0.00026 $ &  $-1$ & 71.22/84 & 16.92 & 14.72 \\
\hline
A2   & $0.674\pm 0.005$& $0.02232\pm0.00014 $&$0.965\pm0.004$& $0.303\pm0.005$ &$0.00150\pm 0.00035 $& $-1$  & 70.27/84 & 17.87 & 15.67\\
\hline
G1 & $0.670\pm 0.006$& $0.02236\pm0.00014 $&$0.967\pm0.004$& $0.302\pm0.005$ &$0.00114\pm 0.00027 $& $-1$  &  71.19/84 & 16.95 & 14.75\\
\hline
G2  & $0.670\pm 0.006$& $0.02234\pm0.00014 $&$0.966\pm0.004$& $0.303\pm0.005$ &$0.00136\pm 0.00032 $& $-1$  &  70.68/84 & 17.46 & 15.26\\
\hline
\end{tabular}}
\end{center}
\label{tableFit}
\begin{scriptsize}
\tablecomments{The best-fit values for the $\CC$CDM, XCDM and the RVM's, including their statistical  significance ($\chi^2$-test and Akaike and Bayesian information criteria, AIC and BIC, see the text). The large and positive values of $\Delta$AIC and $\Delta$BIC strongly favor the dynamical DE options (RVM's and XCDM) against the $\CC$CDM (see text). We use $90$ data points in our fit, to wit: $31$ points from the JLA sample of SNIa, $11$ from BAO, $30$  from $H(z)$, $13$ from linear growth, $1$ from BBN, and $4$ from CMB (see S1-S7 in the text for references). In the XCDM model the EoS parameter $\omega$ is left free, whereas for the RVM's and $\CC$CDM is fixed at $-1$.  The specific RVM fitting parameter is $\nueff$, see Eq.\,(\ref{eq:xixip}) and the text. For G1 and A1 models, $\nueff=\nu$. The remaining parameters are the standard ones ($h,\omega_b,n_s,\Omega_m$).  The quoted number of degrees of freedom ($dof$) is equal to the number of data points minus the number of independent fitting parameters ($5$ for the $\CC$CDM, $6$ for the RVM's and the XCDM. The normalization parameter M introduced in the SNIa sector of the analysis is also left free in the fit, cf. \cite{BetouleJLA}, but it is not listed in the table). For the CMB data we have used the marginalized mean values and standard deviation for the parameters of the compressed likelihood for Planck 2015 TT,TE,EE + lowP data from \cite{Huang}, which provide tighter constraints to the CMB distance priors than those presented in \cite{PlanckDE2015}.}
\end{scriptsize}
\end{table*}
\begin{table*}
\caption{Best-fit values for the various vacuum models and the XCDM using the Planck 2015 results and removing the BAO and LSS data from WiggleZ}
\begin{center}
\resizebox{1\textwidth}{!}{
\begin{tabular}{| c | c |c | c | c | c | c| c | c | c |}
\multicolumn{1}{c}{Model} &  \multicolumn{1}{c}{$h$} &  \multicolumn{1}{c}{$\omega_b= \Omega_b h^2$} & \multicolumn{1}{c}{{\small$n_s$}}  &  \multicolumn{1}{c}{$\Omega_m$}&  \multicolumn{1}{c}{{\small$\nu_{eff}$}}  & \multicolumn{1}{c}{$\omega$}  &
\multicolumn{1}{c}{$\chi^2_{\rm min}/dof$} & \multicolumn{1}{c}{$\Delta{\rm AIC}$} & \multicolumn{1}{c}{$\Delta{\rm BIC}$}\vspace{0.5mm}
\\\hline
{\small $\Lambda$CDM} & $0.692\pm 0.004$ & $0.02254\pm 0.00013$ &$0.975\pm 0.004$& $0.295\pm 0.004$ & - & $-1$   & 86.11/78 & - & -\\
\hline
XCDM  &  $0.671\pm 0.007$& $0.02263\pm 0.00014 $&$0.976\pm 0.004$& $0.312\pm 0.007$& - & $-0.920\pm0.022$  & 73.01/77 & 10.78 & 8.67 \\
\hline
A1  & $0.670\pm 0.007$& $0.02238\pm0.00014 $&$0.967\pm0.004$& $0.302\pm0.005$ &$0.00110\pm 0.00028 $ & $-1$  & 69.40/77 & 14.39 & 12.27 \\
\hline
A2   & $0.674\pm 0.005$& $0.02233\pm0.00014 $&$0.966\pm0.004$& $0.302\pm0.005$ &$0.00152\pm 0.00037 $& $-1$  & 68.38/77 & 15.41 & 13.29\\
\hline
G1 & $0.671\pm 0.006$& $0.02237\pm0.00014 $&$0.967\pm0.004$& $0.302\pm0.005$ &$0.00115\pm 0.00029 $& $-1$  &  69.37/77 & 14.42 & 12.30\\
\hline
G2  & $0.670\pm 0.006$& $0.02235\pm0.00014 $&$0.966\pm0.004$& $0.302\pm0.005$ &$0.00138\pm 0.00034 $& $-1$  &  68.82/77 & 14.97 & 12.85\\
\hline
\end{tabular}}
\end{center}
\begin{scriptsize}
\tablecomments{Same as in Table 1, but excluding from our analysis the BAO and LSS data from WiggleZ, see point S5) in the text.}
\end{scriptsize}
\label{tableFit2}
\end{table*}


\section{Two basic types of RVM's}\label{sect:RVMs}
In an expanding Universe we may expect that the vacuum energy density and the gravitational coupling are functions of the cosmic time through the Hubble rate, thence $\rL=\rL(H(t))$ and $G=G(H(t))$. Adopting the canonical equation of state $p_\Lambda=-\rho_\Lambda(H)$ also for the dynamical vacuum,
the corresponding field equations in the
Friedmann-Lema\^\i tre-Robertson-Walker (FLRW) metric in flat space become formally identical to those
with strictly constant $G$ and $\CC$:
\begin{eqnarray}\label{eq:FriedmannEq}
&&3H^2=8\pi\,G(H)\,(\rho_m+\rR+\rho_\Lambda(H))\\
&&3H^2+2\dot{H}=-8\pi\,G(H)\,(p_r-\rho_\Lambda(H))\,. \label{eq:PressureEq}\,
\end{eqnarray}
The equations of state for the densities of relativistic ($\rho_r$) and dust matter ($\rho_m$) read
$p_r=(1/3)\rho_r$ and $p_m=0$, respectively.  Consider now the characteristic RVM structure
of the dynamical vacuum energy:
\begin{eqnarray}\label{eq:rhoL}
\rL(H;\nu,\alpha)=\frac{3}{8\pi G}\left(c_0+\nu H^2+\frac{2}{3}\alpha\,\dH\right)+{\cal O}(H^4)\,,\label{eq:rL}
\end{eqnarray}
where $G$ can be constant or a function $G=G(H;\nu,\alpha)$ depending on the particular model. The above expression is the form that has been suggested in the literature from the quantum corrections of QFT in curved spacetime (cf. \,\cite{JSPRev2013,SolGom2015} and references therein). The terms  with higher powers of the Hubble rate have recently been used to describe inflation,
see e.g. \cite{LimBasSol131516} and \cite{Sola2015}, but these terms play no role at present and will be hereafter omitted. The coefficients $\nu$ and $\alpha$ have been defined dimensionless. They are responsible for the running of $\rL(H)$ and $G(H)$, and so for $\nu=\alpha=0$ we recover the $\CC$CDM, with $\rL$ and $G$ constants. The values of $\nu$ and $\alpha$ are naturally small in this context since they can be related to the $\beta$-functions of the running. An estimate in QFT indicates that they are of order $10^{-3}$ at most \citep{Fossil07}, but here we will treat them as free parameters of the RVM and hence we shall determine them phenomenologically by fitting the model to observations. As previously indicated, a simple Lagrangian language for these models that is comparable to the scalar field DE description may not be possible, as suggested by attempts involving the anomaly-induced action\,\citep{Fossil07,JSPRev2013}.

Two types of RVM will be considered here: i) type-G models,  when matter is conserved and the running of $\rL(H)$ is compatible with the Bianchi identity at the expense of a (calculable) running of $G$; ii) type-$A$ models, in contrast, denote those with $G=$const. in which the running of $\rL$ must be accompanied with a (calculable) anomalous conservation law of matter. Both situations are described by the generalized local conservation equation $\nabla^{\mu}\left(G\,\tilde{T}_{\mu\nu}\right)=0$, where $\tilde{T}_{\mu\nu}=T_{\mu\nu}+\rL\,g_{\mu\nu}$ is the total energy-momentum tensor involving both matter and vacuum energy. In the FLRW metric, and summing over all energy components, we find
\begin{equation}\label{BianchiGeneral}
\frac{d}{dt}\,\left[G(\rho_m+\rho_r+\rL)\right]+3\,G\,H\,\sum_{i=m,r}(\rho_i+p_i)=0\,.
\end{equation}
If $G$ and $\rL$ are both constants, we recover the canonical  conservation law $\dot{\rho}_m+\dot{\rho}_r+3H\rho_m+4H\rho_r=0$ for the combined system of matter and radiation.
For type-G models Eq.\,(\ref{BianchiGeneral}) boils down to $\dG(\rma+\rr+\rL)+G\drL=0$
since $\dot\rho_m+3H\rmr=0$ and $\dot\rho_r+4H\rR=0$ for separated conservation of matter and radiation, as usually assumed.  Mixed type of RVM scenarios are possible, but will not be considered here.

We can solve analytically the type-G and type-A models by inserting equation \,(\ref{eq:rL}) into (\ref{eq:FriedmannEq}) and (\ref{eq:PressureEq}), or using one of the latter two and the corresponding conservation law (\ref{BianchiGeneral}). It is convenient to perform the  integration using the scale factor $a(t)$  rather than the cosmic time. For type-G models the full expression for the  Hubble function normalized to its current value, $E(a)=H(a)/H_0$, can be found to be
\begin{eqnarray}\label{eq:DifEqH} &&\left.E^2(a)\right|_{\rm type-G}=1+\left(\frac{\Omega_m}{\xi}+\frac{\Omega_r}{\xi^\prime}\right)\nonumber\\
&&\times\left[-1+a^{-4\xi^\prime}\left(\frac{a\xi^\prime+\xi\Omega_r/\Omega_m}{\xi^\prime+\xi\Omega_r/\Omega_m}\right)^{\frac{\xi^\prime}{1-\alpha}}\right]\,, \end{eqnarray}
where $\Omega_{i}=\rho_{i0}/\rho_{c0}$ are the current cosmological parameters for matter and radiation, and we have defined
\be\label{eq:xixip}
\xi=\frac{1-\nu}{1-\alpha}\equiv 1-\nueff\,,\ \ \  \xi^\prime=\frac{1-\nu}{1-\frac{4}{3}\alpha}\equiv 1-\nueffp\,.
\ee
\begin{figure*}
\centering
\includegraphics[angle=0,width=0.9\linewidth]{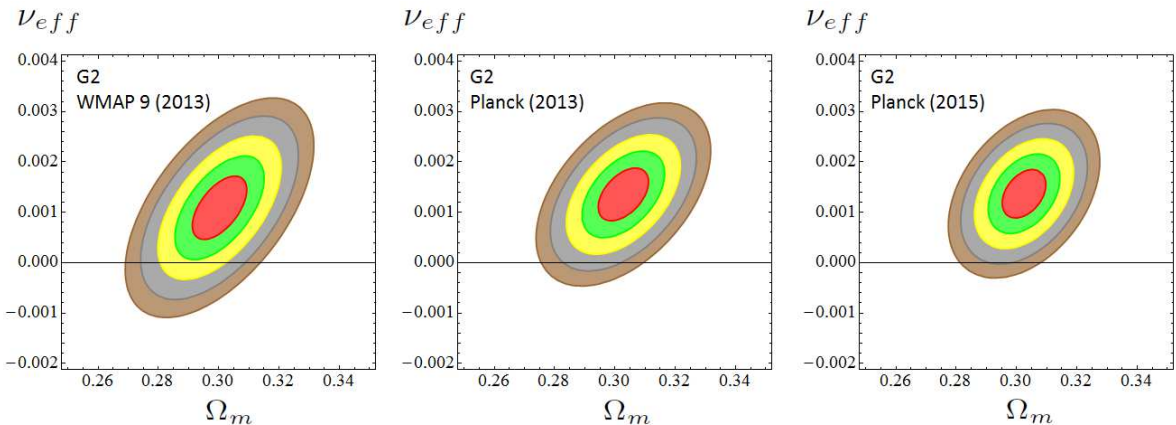}
\caption{\label{fig:G2Evolution}%
\scriptsize Likelihood contours in the $(\Omega_m,\nueff)$ plane for the values $-2\ln\mathcal{L}/\mathcal{L}_{max}=2.30$, $6.18, 11.81$, $19.33$, $27.65$ (corresponding to 1$\sigma$, 2$\sigma$, 3$\sigma$, 4$\sigma$ and 5$\sigma$ c.l.) after marginalizing over the rest of the fitting parameters indicated in Table 1. We display the progression of the contour plots obtained for model G2 using the 90 data points on SNIa+BAO+$H(z)$+LSS+BBN+CMB, as we evolve from the high precision CMB data from WMAP9, Planck 2013 and Planck 2015 -- see text, point S7). In the sequence, the prediction of the concordance  model ($\nueff=0$) appears increasingly more disfavored, at an exclusion c.l. that ranges from  $\sim 2\sigma$ (for WMAP9), $\sim 3.5\sigma$ (for Planck 2013) and up to  $4\sigma$ (for Planck 2015). Subsequent marginalization over $\Omega_m$ increases slightly the c.l. and renders the fitting values indicated in Table 1, which reach a statistical significance of $4.2\sigma$ for all the RVM's. Using numerical integration we can estimate that $\sim99.81\%$ of the area of the $4\sigma$ contour for Planck 2015 satisfies $\nueff>0$. We also estimate that $\sim95.47\%$ of the $5\sigma$ region also satisfies $\nueff>0$. The corresponding AIC and BIC criteria (cf. Table 1) consistently imply a very strong support to the RVM's against the $\CC$CDM.
}
\end{figure*}

Note that $E(1)=1$, as it should. Moreover, for  $\xi,\xi'\to 1$ (i.e. $|\nu,\alpha|\ll 1$)  $\nueff\simeq \nu-\alpha$ and $\nueffp\simeq \nu-(4/3)\alpha$.
In the radiation-dominated epoch, the leading behavior of Eq.\,(\ref{eq:DifEqH}) is $\sim \ORo\,a^{-4\xi'}$, while in the matter-dominated epoch is $\sim \Omo\,a^{-3\xi}$. Furthermore, for $\nu,\alpha\to 0$,  $E^2(a)\to 1+\Omo\,(a^{-3}-1)+\ORo(a^{-4}-1)$. This is the $\CC$CDM form, as expected in that limit. Note
that the following constraint applies among the parameters:
$c_0=H_0^2\left[\OLo-\nu+\alpha\left(\Omo+\frac43\,\ORo\right)\right]$,
as the vacuum energy density $\rL(H)$ must reproduce the current
value $\rLo$ for $H=H_0$, using $\Omo+\ORo+\OLo=1$.  The explicit scale factor dependence of the vacuum energy density, i.e. $\rL=\rL(a)$,
ensues upon inserting  \eqref{eq:DifEqH} into \eqref{eq:rL}. In addition, since the matter is conserved for type-G models, we can use the obtained expression for $\rL(a)$ to also infer the explicit form for $G=G(a)$ from (\ref{eq:FriedmannEq}).
We refrain from writing out these cumbersome expressions and we limit ourselves to quote some simplified forms. For instance, the expression for $\rL(a)$ when we can neglect the radiation contribution is simple enough:
\be\label{eq:RhoLNR} \rL(a)=\rho_{c0}\,
a^{-3}\left[a^{3\xi}+\frac{\Omo}{\xi}(1-\xi-a^{3\xi})\right]\,,
\ee
where  $\rho_{c0}=3H_0^2/8\pi\,G_0$ is the current critical density and $G_0\equiv G(a=1)$ is the current value of the gravitational coupling. Quite obviously for $\xi=1$ we recover the $\CC$CDM form: $\rL=\rho_{c0}(1-\Omo)=\rho_{c0}\OLo=$const. As for the gravitational coupling, it
evolves logarithmically with the scale factor and hence changes very slowly\footnote{This is a welcome feature already expected in particular  realizations of type-G models in QFT in curved spacetime\,\citep{Fossil07,JSPRev2013}. See also \cite{Grande2011}.}. It suffices to say that it behaves as
\begin{equation}\label{Gafunction}
G(a)=G_0\,a^{4(1-\xi')}\,f(a)\simeq G_{0}(1+4\nueffp\,\ln\,a)\,f(a)\,,
\end{equation}
where $f(a)=f(a;\Omo,\ORo; \nu,\alpha)$ is a smooth function of the scale factor. We can dispense with the full expression here, but let us mention that $f(a)$ tends to one at present irrespective of the values of the various parameters $\Omo,\ORo,\nu,\alpha$ involved in it; and $f(a)\to1$ in the remote past ($a\to 0$) for $\nu,\alpha\to 0$ (i.e. $\xi,\xi'\to 1$). As expected, $G(a)\to G_0$ for $a\to 1$, and $G(a)$ has a logarithmic evolution for $\nueffp\neq 0$.
Notice that the limit $a\to 0$ is  relevant for the BBN (Big Bang Nucleosynthesis) epoch and therefore $G(a)$ should not depart too much from $G_0$ according to the usual bounds on BBN. We shall carefully incorporate this restriction in our analysis of the RVM models, see later on.

Next we quote the solution for type-A models. As indicated, in this case we have an anomalous matter conservation law. Integrating (\ref{BianchiGeneral}) for $G=$const. and using \eqref{eq:rL} in it one finds $\rho_t(a)\equiv\rho_m(a)+\rho_r(a)=\rho_{m0}a^{-3\xi}+\rho_{r0}a^{-4\xi'}$. We have assumed, as usual, that there is no exchange of energy between the relativistic and non-relativistic components. The standard expressions for matter and radiation energy densities are  recovered for $\xi,\xi'\to 1$. The normalized Hubble function for type-A models is simpler than for type-G ones. The full expression including both matter and radiation reads:
\begin{equation}\label{eq:HubbleA}
\left.E^2(a)\right|_{\rm type-A}=1+\frac{\Omega_m}{\xi}\left(a^{-3\xi}-1\right)+\frac{\Omega_r}{\xi^\prime}\left(a^{-4\xi^\prime}-1\right)\,.
\end{equation}
From it and the found expression for $\rho_t(a)$ we can immediately derive the corresponding $\rL(a)$:
\begin{equation}\label{rLaTypaA}
\rho_\CC(a)=\rLo+\rmo(\xi^{-1}-1)(a^{-3\xi}-1)+\rRo(\xi^{\prime-1}-1)(a^{-4\xi^\prime}-1)\,.
\end{equation}
Once more for $\nu,\alpha\to 0$ (i.e. $\xi,\xi'\to 1$) we recover the $\CC$CDM case, as easily checked. In particular one finds $\rL\to\rLo=$const. in this limit.

\begin{figure*}
\centering
\includegraphics[angle=0,width=0.9\linewidth]{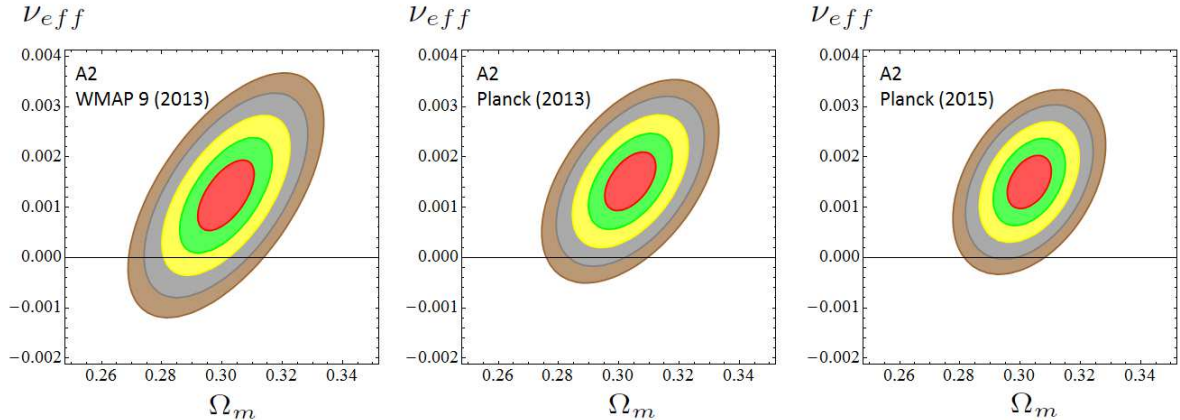}
\caption{\label{fig:A1Evolution}%
\scriptsize As in Fig.\,1, but for model A2. Again we see that the contours tend to migrate to the $\nueff>0$ half plane as we evolve from WMAP9 to Planck 2013 and Planck 2015 data. Using the same method as in Fig.\,1, we find that $\sim99.82\%$ of the area of the $4\sigma$ contour for Planck 2015 (and  $\sim95.49\%$ of the corresponding $5\sigma$ region) satisfies $\nueff>0$.  The $\CC$CDM  becomes once more excluded at $\sim 4\sigma$ c.l.  ({cf. Table 1 for Planck 2015}).
}
\end{figure*}
\begin{figure}
\centering
\includegraphics[angle=0,width=0.65\linewidth]{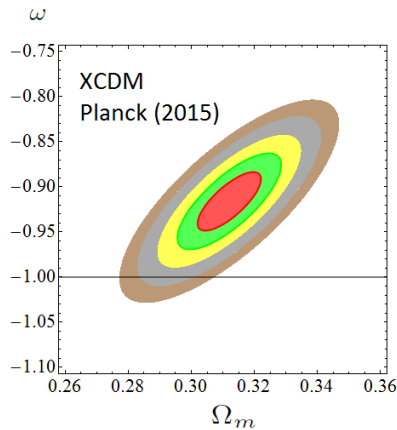}
\caption{\label{fig:XCDMEvolution}%
\scriptsize As in Fig.\,1 and 2, but for model XCDM and using Planck 2015 data. The $\CC$CDM is excluded at $\sim 4\sigma$ c.l.  ({cf. Table 1}).
}
\end{figure}

\section{Fitting the vacuum models to the data}\label{sect:Fit}

In order to better handle the possibilities offered by the type-G and type-A models as to their dependence on the two specific vacuum parameters $\nu,\alpha$, we shall refer to model G1 (resp. A1) when we address type-G (resp. type-A) models with $\alpha=0$ in Eq.\,(\ref{eq:rL}). In these cases $\nueff=\nu$.  When, instead, $\alpha\neq 0$ we shall indicate them by G2 and A2, respectively. This classification scheme is used in Tables 1-2 and 5-7, and in Figs. 1-6. In the tables we are including also the XCDM (cf. Section 4) and the $\CC$CDM.

To this end, we fit the various models to the wealth of cosmological data compiled from distant type Ia supernovae (SNIa), baryonic acoustic oscillations (BAO's), the known values of the Hubble parameter at different redshift points, $H(z_i)$, the large scale structure (LSS) formation data encoded in $f(z_i)\sigma_8(z_i)$, the BBN bound on the Hubble rate, and, finally, the CMB distance priors from WMAP and Planck, with the corresponding correlation matrices in all the indicated cases. Specifically, we have used $90$ data points (in some cases involving compressed data) from $7$ different sources S1-S7, to wit:
\begin{table}
 \caption{Compilation of $H(z)$ data points}
\begin{center}
\begin{tabular}{| c | c | c |}
\multicolumn{1}{c}{$z$} &  \multicolumn{1}{c}{$H(z)$} & \multicolumn{1}{c}{{\small References}}
\\\hline
$0.07$ & $69.0\pm 19.6$ & \cite{Zhang}
\\\hline
$0.09$ & $69.0\pm 12.0$ & \cite{Jimenez}
\\\hline
$0.12$ & $68.6\pm 26.2$ & \cite{Zhang}
\\\hline
$0.17$ & $83.0\pm 8.0$ & \cite{Simon}
\\\hline
$0.1791$ & $75.0\pm 4.0$ & \cite{Moresco2012}
\\\hline
$0.1993$ & $75.0\pm 5.0$ & \cite{Moresco2012}
\\\hline
$0.2$ & $72.9\pm 29.6$ & \cite{Zhang}
\\\hline
$0.27$ & $77.0\pm 14.0$ & \cite{Simon}
\\\hline
$0.28$ & $88.8\pm 36.6$ & \cite{Zhang}
\\\hline
$0.3519$ & $83.0\pm 14.0$ & \cite{Moresco2012}
\\\hline
$0.3802$ & $83.0\pm 13.5$ & \cite{Moresco2016}
\\\hline
$0.4$ & $95.0\pm 17.0$ & \cite{Simon}
\\\hline
$0.4004$ & $77.0\pm 10.2$ & \cite{Moresco2016}
\\\hline
$0.4247$ & $87.1\pm 11.2$ & \cite{Moresco2016}
\\\hline
$0.4497$ & $92.8\pm 12.9$ & \cite{Moresco2016}
\\\hline
$0.4783$ & $80.9\pm 9.0$ & \cite{Moresco2016}
\\\hline
$0.48$ & $97.0\pm 62.0$ & \cite{Stern}
\\\hline
$0.5929$ & $104.0\pm 13.0$ & \cite{Moresco2012}
\\\hline
$0.6797$ & $92.0\pm 8.0$ & \cite{Moresco2012}
\\\hline
$0.7812$ & $105.0\pm 12.0$ & \cite{Moresco2012}
\\\hline
$0.8754$ & $125.0\pm 17.0$ & \cite{Moresco2012}
\\\hline
$0.88$ & $90.0\pm 40.0$ & \cite{Stern}
\\\hline
$0.9$ & $117.0\pm 23.0$ & \cite{Simon}
\\\hline
$1.037$ & $154.0\pm 20.0$ & \cite{Moresco2012}
\\\hline
$1.3$ & $168.0\pm 17.0$ & \cite{Simon}
\\\hline
$1.363$ & $160.0\pm 33.6$ & \cite{Moresco2015}
\\\hline
$1.43$ & $177.0\pm 18.0$ & \cite{Simon}
\\\hline
$1.53$ & $140.0\pm 14.0$ & \cite{Simon}
\\\hline
$1.75$ & $202.0\pm 40.0$ & \cite{Simon}
\\\hline
$1.965$ & $186.5\pm 50.4$ & \cite{Moresco2015}
\\\hline
\end{tabular}
\end{center}
\label{compilationH}
\begin{scriptsize}
\tablecomments{Current published values of $H(z)$ in units [km/s/Mpc] obtained using the differential-age technique (see the quoted references and point S4 in the text).}
\end{scriptsize}
\end{table}
\begin{table}
 \caption{Compilation of $f(z)\sigma_8(z)$ data points}
\begin{center}
\begin{tabular}{| c | c |c | c |}
\multicolumn{1}{c}{Survey} &  \multicolumn{1}{c}{$z$} &  \multicolumn{1}{c}{$f(z)\sigma_8(z)$} & \multicolumn{1}{c}{{\small References}}
\\\hline
6dFGS & $0.067$ & $0.423\pm 0.055$ & \cite{Beutler2012}
\\\hline
SDSS-DR7 & $0.10$ & $0.37\pm 0.13$ & \cite{Feix}
\\\hline
GAMA & $0.18$ & $0.29\pm 0.10$ & \cite{Simpson}
\\ \cline{2-4}& $0.38$ & $0.44\pm0.06$ & \cite{Blake2013}
\\\hline
DR12 BOSS & $0.32$ & $0.427\pm 0.052$  & \cite{GilMarin2}\\ \cline{2-3}
 & $0.57$ & $0.426\pm 0.023$ & \\\hline
 WiggleZ & $0.22$ & $0.42\pm 0.07$ & \cite{Blake2011fs8} \tabularnewline
\cline{2-3} & $0.41$ & $0.45\pm0.04$ & \tabularnewline
\cline{2-3} & $0.60$ & $0.43\pm0.04$ & \tabularnewline
\cline{2-3} & $0.78$ & $0.38\pm0.04$ &
\\\hline
2MTF & $0.02$ & $0.34\pm 0.04$ & \cite{Springob}
\\\hline
VIPERS & $0.7$ & $0.380\pm0.065$ & \cite{Granett}
\\\hline
VVDS & $0.77$ & $0.49\pm0.18$ & \cite{Guzzo}\tabularnewline
 & & &\cite{Percival08}
\\\hline
 \end{tabular}
\end{center}
\label{compilationLSS}
\begin{scriptsize}
\tablecomments{Current published values of $f(z)\sigma_8(z)$. See the text, S5).}
\end{scriptsize}
\end{table}

S1) The SNIa data points from the SDSS-II/SNLS3 Joint Light-curve Analysis (JLA) \citep{BetouleJLA}. We have used the $31$ binned distance modulus fitted to the JLA sample and the compressed form of the likelihood with the corresponding covariance matrix.

S2) 5 points on the isotropic BAO estimator $r_s(z_d)/D_V(z_i)$: $z=0.106$ \citep{Beutler2011}, $z=0.15$\, \citep{Ross}, $z_i=0.44, 0.6, 0.73$ \citep{Kazin2014}, with the correlations between the last 3 points.

S3) 6 data points on anisotropic BAO estimators: 4 of them on  $D_A(z_i)/r_s(z_d)$ and $H(z_i)r_s(z_d)$ at $z_i=0.32, 0.57$, for the LOWZ and CMASS samples, respectively. These data are taken from \cite{GilMarin2}, based on the Redshift-Space Distortions (RSD) measurements of the power spectrum combined with the bispectrum, and the BAO post-reconstruction analysis of the power spectrum (cf. Table 5 of that reference), including the correlations among these data encoded in the provided covariance matrices. We also use  2 data points  based on  $D_A(z_i)/r_s(z_d)$ and $D_H(z_i)/r_s(z_d)$ at $z=2.34$, from the combined LyaF analysis \cite{Delubac}. The correlation coefficient among these 2 points are taken from \cite{Aubourg} (cf. Table II of that reference). We also take into account the correlations among the  BAO data and the corresponding $f\sigma_8$ data of \cite{GilMarin2} -- see S5) below and Table 4.

S4) $30$ data points on $H(z_i)$ at different redshifts, listed in Table 3. We use only $H(z_i)$ values obtained by the so-called differential-age techniques applied to passively evolving galaxies. These values are  uncorrelated with the BAO data points. See also \cite{FarooqRatra2013}, \cite{SahniShafielooStarobinsky}, \cite{Ding}, \cite{Zheng} and \cite{ChenKumarRatra2016}, where the authors make only use of Hubble parameter data in their analyses. We find, however, indispensable to take into account the remaining data sets to derive our conclusions on dynamical vacuum, specially the BAO, LSS and CMB observations. This fact can also be verified quite evidently in {Figures 5-6}, to which we shall turn our attention in Section 4.

S5) $f(z)\sigma_8(z)$: 13 points. These are referred to in the text as LSS (large scale structure formation). {The actual fitting results shown in Table 1 make use of the LSS data listed in Table 4, in which we have carefully avoided possible correlations among them (see below). Let us mention that although we are aware of the existence of other LSS data points in the literature concerning some of the used redshift values in our Table 4 -- cf. e.g. \cite{Percival2004}; \cite{Turnbull2012,HudsonTurnbull2013}; \cite{Johnson2014} -- we have explicitly checked that their inclusion or not in our numerical fits has no significant impact on the main result of our paper, that is to say, it does not affect the attained $\gtrsim4\sigma$ level of evidence in favor of the RVM's. This result is definitely secured in both cases, but we have naturally presented our final results sticking to the most updated data. }

The following observation is also in order. We have included both the WiggleZ and the CMASS data sets in our analysis. We are aware that there exists some overlap region between the CMASS and WiggleZ galaxy samples. But the two surveys have
been produced independently and the studies on the existing correlations among these observational results \citep{Beutler2016,Marin2016} show that the correlation is small. The overlap region of the CMASS and WiggleZ galaxy samples is actually not among the galaxies that the two surveys pick up, but between the region of the sky they explore. Moreover, despite almost all the WiggleZ region (5/6 parts of it) is inside the CMASS one, it only takes a very small fraction of the whole sky region covered by CMASS, since the latter is much larger than the WiggleZ one (see, e.g. Figure 1 in \cite{Beutler2016}). In this paper, the authors are able to quantify the correlation degree among the BAO constraints in CMASS and WiggleZ, and they conclude that it is less than 4\%. Therefore, we find it justified to include the WiggleZ data in the main table of results of our analysis (Table 1), but we provide also the fitting results that are obtained when we remove the WiggleZ data points from the BAO and $f(z)\sigma_8(z)$ data sets (see Table 2). The difference is small and the central values of the fitting parameters and their uncertainties remain intact. Thus the statistical significance of Tables 1 and 2 is the same.

S6) BBN:  we have imposed the average bound on the possible variation of the BBN speed-up factor, defined as the ratio of the expansion rate predicted in a given model versus that of the  $\CC$CDM model at the BBN epoch ($z\sim 10^9$). This amounts to the limit $|\Delta H^2/H_\Lambda^2|<10\%$ \,\citep{Uzan}.

S7) CMB distance priors:  $R$ (shift parameter) and $\ell_a$ (acoustic length) and their correlations  with $(\omega_b,n_s)$. For WMAP9 and Planck 2013 data we used the covariance matrix from the analysis of \cite{WangWang}, while for Planck 2015 data those of \cite{Huang}. Our fitting results for the last case are recorded in all our tables (except in Table 5 where we test our fit in the absence of CMB distance priors $R$ and $\ell_a$). We display the final contour plots for all the cases, see Figs. 1-2. Let us point out that in the case of the Planck 2015 data we have checked that very similar results ensue for all models if we use the alternative CMB covariance matrix from \cite{PlanckDE2015}. We have, however, chosen to explicitly present the case based on  \cite{Huang} since it uses the more complete compressed likelihood analysis for Planck 2015 TT,TE,EE + lowP data whereas \cite{PlanckDE2015} uses Planck 2015 TT+lowP data only.

Notice that G1 and A1 have one single vacuum parameter ($\nu$) whereas G2 and A2 have two ($\nu,\alpha$). There is nonetheless a natural alignment between $\nu$ and $\alpha$ for general type-G and A models, namely $\alpha=3\nu/4$, as this entails $\xi'=1$ (i.e. $\nueffp=0$) in Eq.\,(\ref{eq:xixip}). Recall that for G2 models we have $G(a)\sim G_0\,a^{4(1-\xi')}$ deep in the radiation epoch, cf. Eq.\,(\ref{Gafunction}), and therefore the condition $\xi'=1$ warrants $G$ to take the same value as the current one, $G=G_0$,  at BBN. For model G1 this is not possible (for $\nu\neq 0$) and we adopt the aforementioned $|\Delta H^2/H_{\CC}^2|<10\%$ bound. We apply the same BBN restrictions to the A1 and A2 models, which have constant $G$. With this setting all the vacuum models contribute only with one single additional parameter as compared to the $\CC$CDM: $\nu$, for G1 and A1; and $\nueff=\nu-\alpha=\nu/4$, for G2 and A2.

For the statistical analysis, we define the joint likelihood function as the product of the likelihoods for all the data sets. Correspondingly, for Gaussian errors the total $\chi^2$ to be minimized reads:
\be
\chi^2_{tot}=\chi^2_{SNIa}+\chi^2_{BAO}+\chi^2_{H}+\chi^2_{f\sigma_8}+\chi^2_{BBN}+\chi^2_{CMB}\,.
\ee
Each one of these terms is defined in the standard way, for some more details see e.g. \cite{GomSolBas2015}, although we should emphasize that here the correlation matrices have been included. The BAO part was split as indicated in S2) and S3) above. Also, in contrast to the previous analysis of \cite{ApJ1}, we did not use here the correlated $Omh^2(z_i,z_j)$ diagnostic for $H(z_i)$ data. Instead, we use
\begin{equation}\label{chi2H}
\chi^{2}_{\rm H}({\bf p})=\sum_{i=1}^{30} \left[ \frac{ H(z_{i},{\bf p})-H_{\rm obs}(z_{i})}
{\sigma_{H,i}} \right]^{2}\,.
\end{equation}
%
As for the linear structure formation data we have computed the density contrast $\delta_m=\delta\rho_m/\rho_m$ for each vacuum model by adapting the cosmic perturbations formalism for type-G and type-A vacuum models.
The matter perturbation, $\delta_m$, obeys a generalized equation which depends on the RVM type. For type-A models it reads (as a differential equation with respect to the cosmic time)
\begin{equation}\label{diffeqD}
\ddot{\delta}_m+\left(2H+\Psi\right)\,\dot{\delta}_m-\left(4\pi
G\rmr-2H\Psi-\dot{\Psi}\right)\,\delta_m=0\,,
\end{equation}
where $\Psi\equiv-\frac{\dot{\rho}_{\CC}}{\rmr}$. For $\rL=$const. we have $\Psi=0$ and Eq.\,(\ref{diffeqD}) reduces to the $\CC$CDM form\,\footnote{For details on these equations, confer the comprehensive works \cite{GomSolBas2015}, \cite{GomSolMNRAS} and \cite{GoSolKarim2015}.}. For type-G models the matter perturbation equation is explicitly given in \cite{ApJ1}. From here we can derive the weighted linear growth $f(z)\sigma_8(z)$, where $f(z)=d\ln{\delta_m}/d\ln{a}$ is the growth factor and $\sigma_8(z)$ is the rms mass fluctuation amplitude on scales of $R_8=8\,h^{-1}$ Mpc at redshift $z$. {It is computed from}
\begin{equation}
\begin{small}\sigma_8^2(z)=\frac{\delta_m^2(z)}{2\pi^2}\int_{0}^{\infty}k^2\,P(k,\vec{p})\,W^2(kR_8)\,dk\,,\label{s88generalNN}
\end{small}\end{equation}
with $W$ a top-hat smoothing function (see e.g. \cite{GomSolBas2015} for
details). The linear matter power spectrum reads $P(k,\vec{p})=P_0k^{n_s}T^2(k,\vec{p})$, where $\vec{p}=(h,\omega_b,n_s,\Omega_m,\nueff)$ is the fitting vector for the vacuum models we are analyzing (including the $\Lambda$CDM, for which $\nueff=0$ of course), and $T(\vec{p},k)$ is the transfer function, which we take from\,\cite{Bardeen}, {upon introducing the baryon density effects through the modified shape parameter $\Gamma$\,\citep{PeacockDodds,Sugiyama}. We have also explicitly checked that the use of the effective shape of the transfer function provided in \cite{EisensteinHu1998} does not produce any change in our results.}

{The expression (\ref{s88generalNN}) at $z=0$ allows us to write $\sigma_8(0)$ in terms of the power spectrum normalization factor $P_0$ and the primary parameters that enter our fit for each model (cf. Table 1). We fix $P_0$ from}
\begin{equation}
\begin{small}P_0=2\pi^2\frac{\sigma_{8,\Lambda}^2}{\delta^2_{m,\Lambda}(0)}\left[\int_0^\infty k^{2+n_{s,\Lambda}}T^2(k,\vec{p}_\Lambda)W^2(kR_{8,\Lambda})dk\right]^{-1}\,,\label{P0}
\end{small}\end{equation}
{in which we have introduced the vector of fiducial parameters $\vec{p}_\CC=(h_{\CC},\omega_{b,\CC},n_{s,\CC},\Omega_{m,\CC},0)$. This vector is defined in analogy with the fitting vector introduced before, but all its parameters are fixed and taken to be equal to those from the Planck 2015 TT,TE,EE+lowP+lensing analysis\,\citep{Planck2015} with $\nueff=0$. The fiducial parameter $\sigma_{8,\Lambda}$ is also taken from the aforementioned Planck 2015 data.  However, $\delta_{m,\Lambda}(0)$ in (\ref{P0}) is computable: it is the value of $\delta_m(z=0)$ obtained from solving the perturbation equation of the $\CC$CDM using the mentioned fiducial values of the other parameters. Finally, from $\sigma_8(z) = \sigma_8(0)\delta_m(z)/\delta_m(0)$ and plugging \eqref{P0} in \eqref{s88generalNN} one finds:}
\begin{equation}
\begin{small}\sigma_{\rm 8}(z)=\sigma_{8, \Lambda}
\frac{\delta_m(z)}{\delta_{m,\CC}(0)}
\left[\frac{\int_{0}^{\infty} k^{2+n_s} T^{2}(k,\vec{p})
W^2(kR_{8}) dk} {\int_{0}^{\infty} k^{2+n_{s,\CC}} T^{2}(k,\vec{p}_\Lambda) W^2(kR_{8,\Lambda}) dk}
\right]^{1/2}\,.\label{s88general}
\end{small}\end{equation}
{Computing next the weighted linear growth rate  $f(z)\sigma_8(z)$ for each model under consideration, including the $\CC$CDM, all models become normalized to the same fiducial model defined above. The results for $f(z)\sigma_8(z)$ in the various cases are displayed in Fig.\,4 together with the LSS data measurements (cf. Table 4). We will further comment on these results in the next section.}

\begin{figure}
\centering
\includegraphics[angle=0,width=0.9\linewidth]{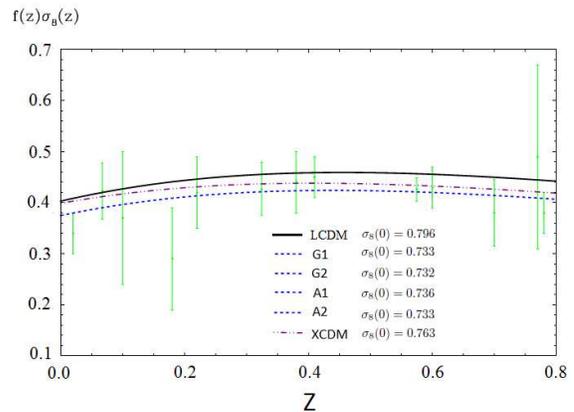}
\caption{\label{fig:fsigma8}%
\scriptsize {The $f(z)\sigma_8(z)$ data (Table 4) and the predicted curves by the RVM's, XCDM and the $\CC$CDM, using the best-fit values in Table 1. Shown are also the values of $\sigma_8(0)$ that we obtain for all the models. The theoretical prediction of all the RVM's are visually indistinguishable and they have been plotted using the same (blue) dashed curve.}
}
\end{figure}

\section{Discussion}

Table 1 and Figures 1-2 present in a nutshell our main results. We observe that the effective vacuum parameter, $\nueff$, is neatly projected non null and positive for all the RVM's. The presence of this effect can be traced already in the old WMAP9 data (at $\sim 2\sigma$), but as we can see it becomes strengthened at $\sim 3.5\sigma$ c.l. with the  Planck 2013 data  and at $\sim4\sigma$ c.l.  with the Planck 2015 data -- see Figs. 1 and 2. For Planck 2015 data it attains up to $\gtrsim4.2\sigma$ c.l. for all the RVM's after marginalizing over the other fitting parameters.

\begin{table*}
\caption{Best-fit values for the various vacuum models and the XCDM using the Planck 2015 results and removing the $R$-shift parameter and the acoustic length $l_a$}
\begin{center}
\resizebox{1\textwidth}{!}{
\begin{tabular}{| c | c |c | c | c | c | c| c | c | c |}
\multicolumn{1}{c}{Model} &  \multicolumn{1}{c}{$h$} &  \multicolumn{1}{c}{$\omega_b= \Omega_b h^2$} & \multicolumn{1}{c}{{\small$n_s$}}  &  \multicolumn{1}{c}{$\Omega_m$}&  \multicolumn{1}{c}{{\small$\nu_{eff}$}}  & \multicolumn{1}{c}{$\omega$} &
\multicolumn{1}{c}{$\chi^2_{\rm min}/dof$} & \multicolumn{1}{c}{$\Delta{\rm AIC}$} & \multicolumn{1}{c}{$\Delta{\rm BIC}$}\vspace{0.5mm}
\\\hline
$\Lambda$CDM  & $0.679\pm 0.005$ & $0.02241\pm 0.00017$ &$0.968\pm 0.005$& $0.291\pm 0.005$ & - & $-1$ & 68.42/83 & - & - \\
\hline
$X$CDM  & $0.673\pm 0.007$& $0.02241\pm 0.00017 $&$0.968\pm 0.005$& $0.299\pm 0.009$& - & $-0.958\pm0.038$  & 67.21/82  & -1.10 & -3.26 \\
\hline
A1  & $0.679\pm 0.010$& $0.02241\pm0.00017 $&$0.968\pm0.005$& $0.291\pm0.010$ &$-0.00001\pm 0.00079 $ & $-1$  & 68.42/82 & -2.31 & -4.47 \\
\hline
A2   & $0.676\pm 0.009$& $0.02241\pm0.00017 $&$0.968\pm0.005$& $0.295\pm0.014$ &$0.00047\pm 0.00139 $& $-1$  & 68.31/82 & -2.20 & -4.36 \\
\hline
G1 & $0.679\pm 0.009$& $0.02241\pm0.00017 $&$0.968\pm0.005$& $0.291\pm0.010$ &$0.00002\pm 0.00080 $& $-1$  &  68.42/82 & -2.31  &  -4.47\\
\hline
G2  & $0.678\pm 0.012$& $0.02241\pm0.00017 $&$0.968\pm0.005$& $0.291\pm0.013$ &$0.00006\pm 0.00123 $& $-1$  &  68.42/82 & -2.31  &  -4.47\\
\hline
\end{tabular}}
\end{center}
\label{tableFitRla}
\begin{scriptsize}
\tablecomments{Same as in Table 1, but removing both the $R$-shift parameter and the acoustic length $l_a$ from our fitting analysis.}
\end{scriptsize}
\end{table*}
\begin{table*}
\caption{Best-fit values for the various vacuum models and the XCDM using the Planck 2015 results and removing the LSS data set}
\begin{center}
\resizebox{1\textwidth}{!}{
\begin{tabular}{| c | c |c | c | c | c | c| c | c | c |}
\multicolumn{1}{c}{Model} &  \multicolumn{1}{c}{$h$} &  \multicolumn{1}{c}{$\omega_b= \Omega_b h^2$} & \multicolumn{1}{c}{{\small$n_s$}}  &  \multicolumn{1}{c}{$\Omega_m$}&  \multicolumn{1}{c}{{\small$\nu_{eff}$}}  & \multicolumn{1}{c}{$\omega$}  &
\multicolumn{1}{c}{$\chi^2_{\rm min}/dof$} & \multicolumn{1}{c}{$\Delta{\rm AIC}$} & \multicolumn{1}{c}{$\Delta{\rm BIC}$}\vspace{0.5mm}
\\\hline
$\Lambda$CDM  & $0.685\pm 0.004$ & $0.02243\pm 0.00014$ &$0.969\pm 0.004$& $0.304\pm 0.005$ & - & $-1$  & 61.70/72 & - & - \\
\hline
XCDM  & $0.683\pm 0.009$ & $0.02245\pm 0.00015$ &$0.969\pm 0.004$& $0.306\pm 0.008$ & - & $-0.991\pm0.040$  & 61.65/71 & -2.30 & -4.29 \\
\hline
A1  & $0.685\pm 0.010$& $0.02243\pm0.00014 $&$0.969\pm0.004$& $0.304\pm0.005$ &$0.00003\pm 0.00062 $ & $-1$  & 61.70/71 & -2.36 & -4.34 \\
\hline
A2   & $0.684\pm 0.009$& $0.02242\pm0.00016 $&$0.969\pm0.005$& $0.304\pm0.005$ &$0.00010\pm 0.00095 $ & $-1$  & 61.69/71 & -2.35 & -4.33 \\
\hline
G1 & $0.685\pm 0.010$& $0.02243\pm0.00014 $&$0.969\pm0.004$& $0.304\pm0.005$ &$0.00003\pm 0.00065 $ & $-1$  & 61.70/71 & -2.36 & -4.34 \\
\hline
G2  & $0.685\pm 0.010$& $0.02242\pm0.00015 $&$0.969\pm0.004$& $0.304\pm0.005$ &$0.00006\pm 0.00082 $ & $-1$  & 61.70/71 & -2.36 & -4.34 \\
\hline
\end{tabular}}
\end{center}
\label{tableFitLSS}
\begin{scriptsize}
\tablecomments{Same as in Table 1, but removing the points from the LSS data set from our analysis,  i.e. all the 13 points on $f\sigma_8$ .}
\end{scriptsize}
\end{table*}
%

It is also interesting to gauge the dynamical character of the DE by performing a fit to the overall data in terms of the well-known XCDM parametrization, in  which the DE is mimicked through the density $\rho_X(a)=\rho_{X0}\,a^{-3(1+\omega)}$ associated to some generic entity X, which acts as an ersatz for the $\CC$ term; $\rho_{X0}$ being the current energy density value of X and therefore equivalent to $\rho_{\CC 0}$, and  $\omega$ is the (constant) equation of state (EoS) parameter for X. The XCDM trivially boils down to the rigid $\CC$-term for $\omega=-1$, but by leaving $\omega$ free it proves a useful approach to roughly mimic a (non-interactive) DE scalar field with constant EoS. The corresponding fitting results are included in all our tables along with those for the RVM's and the $\CC$CDM. {In Table 1 (our main table) and in Fig. 3, we can see that the best fit value for $\omega$ in the XCDM is $\omega=-0.916\pm0.021$. Remarkably, it departs from $-1$ by precisely $4\sigma$.}

{Obviously, given the significance of the above result it is highly convenient to compare it with previous analyses of the XCDM reported by the Planck and BOSS collaborations. The Planck 2015 value for the EoS parameter of the XCDM reads $\omega = -1.019^{+0.075}_{-0.080}$  \citep{Planck2015} and the BOSS one is $\omega = -0.97\pm 0.05$\,\citep{Aubourg}. These results are perfectly compatible with our own result for $\omega$ shown in Table 1 for the XCDM, but in stark contrast to our result their errors are big enough as to be also fully compatible with the $\CC$CDM value $\omega=-1$. This is, however, not too surprising if we take into account that none of these analyses included LSS data in their fits, as explicitly indicated in their papers\,\footnote{Furthermore, at the time these analyses appeared they could not have used the important LSS and BAO results from \citep{GilMarin2}, i.e. those that we have incorporated as part of our current data set, not even the previous ones from\,\citep{GilMarin1}. The latter also carry a significant part of the dynamical DE signature we have found here, as we have checked.}. In the absence of LSS data we would find a similar situation. In fact, as our Table 6 clearly shows, the removal of the LSS data set in our fit induces a significant increase in the magnitude of the central value of the EoS parameter, as well as the corresponding error. This happens because the higher is $|\omega|$ the higher is the structure formation power predicted by the XCDM, and therefore the closer is such prediction with that of the $\CC$CDM (which is seen to predict too much power as compared to the data, see Fig. 4). In these conditions our analysis renders $\omega = -0.991\pm 0.040$, which is definitely closer to (and therefore compatible with) the central values obtained by Planck and BOSS teams. In addition, this result is now fully compatible with the $\CC$CDM, as in the Planck 2015 and BOSS cases, and all of them are unfavored by the LSS observations. This is consistent with the fact that both information criteria, $\Delta$AIC and $\Delta$BIC, become now slightly negative in Table 6, which reconfirms that if the LSS data are not used the $\CC$CDM performance is  comparable or even better than the other models. So in order to fit the  observed values of $f\sigma_8$, which are generally lower than the predicted ones by the $\CC$CDM, $|\omega|$ should decrease. This is exactly what happens for the XCDM, as well as for the RVM's, when the LSS data are included in our analysis (in combination with the other data, particularly with BAO and CMB data). It is apparent from Fig. 4 that the curves for these models are  then shifted below and hence adapt significantly better to the data points. Correspondingly, the quality of the fits increases dramatically, and this is also borne out by the large and positive values of  $\Delta$AIC and $\Delta$BIC, both above $10$ (cf. Table 1).}

{The above discussion explains why our analysis of the observations through the XCDM is sensitive to the dynamical feature of the DE, whereas the previous results in the literature  are not. It also shows that the size of the effect found with such a parametrization of the DE essentially concurs with the strength of the dynamical vacuum signature found for the RVM's using exactly the same data. This is remarkable, and it was not obvious a priori} since for some of our RVM's (specifically for A1 and A2) there is an interaction between vacuum and matter that triggers an anomalous conservation law, whereas for others (G1 and G2) we do not have such interaction (meaning that matter is conserved in them, thereby following the standard decay laws for relativistic and non-relativistic components). The interaction, when occurs, is however proportional to $\nueff$ and thus is small because the fitted value of $\nueff$ is small. This probably explains why the XCDM can succeed in nailing down the dynamical nature of the DE with a comparable performance. However not all dynamical vacuum models describe the data with the same efficiency, see e.g \,\cite{Salvatelli2014}, \cite{Murgia2016}, \cite{Li2016}. A detailed comparison is made among models similar (but different) from those addressed here in \cite{prl}.  In the XCDM case the departure from the $\CC$CDM takes the fashion of  ``effective quintessence'', whereas for the RVM's it appears as genuine vacuum dynamics. In all cases, however, we find unmistakable signs of DE physics beyond the $\CC$CDM (cf. Table 1), and this is a most important result of our work.

As we have discussed in Section 2, for models A1 and A2 there is an interaction between vacuum and matter. Such interaction is, of course, small because the fitted values of $\nueff$  are small, see Table 1. The obtained values are in the ballpark of $\nueff\sim {\cal O}(10^{-3})$ and therefore this is also the order of magnitude associated to the anomalous conservation law of matter. For example, for the non-relativistic component we have
\begin{equation}
\rho_m(a)=\rho_{m0}a^{-3\xi}=\rho_{m0}a^{-3(1-\nueff)}\,.
\end{equation}
This behavior has been used in the works by \cite{FriSol1,FriSol2} as a possible explanation for the hints on the time variation of the fundamental constants, such as coupling constants and particle masses, frequently considered in the literature. The current observational values for such time variation are actually compatible with the fitted values we have found here. This is an intriguing subject that is currently of high interest in the field, see e.g.\,\cite{Uzan} and \cite{Sola2015editor}.
For models G1 and G2, instead, the role played by $\nueff$ and $\nueffp$ is different. It does not produce any anomaly in the traditional matter conservation law (since matter and radiation are conserved for type-G models), but now it impinges a small (logarithmic) time evolution on $G$ in the fashion sketched in Eq.\,(\ref{Gafunction}). Thus we find, once more, a possible description for the potential variation of the fundamental constants, in this case $G$, along the lines of the above cited works, see also \cite{FriNunSol}. There are, therefore, different phenomenological possibilities to test the RVM's considered here from various points of view.

\begin{table*}
\caption{Best-fit values for the various vacuum models and the XCDM using the Planck 2015 data set}
\begin{center}
\resizebox{1\textwidth}{!}{
\begin{tabular}{| c | c |c | c | c | c | c| c | c | c |}
\multicolumn{1}{c}{Model} &  \multicolumn{1}{c}{$h$} &  \multicolumn{1}{c}{$\omega_b= \Omega_b h^2$} & \multicolumn{1}{c}{{\small$n_s$}}  &  \multicolumn{1}{c}{$\Omega_m$}&  \multicolumn{1}{c}{{\small$\nu_{eff}$}}  & \multicolumn{1}{c}{$\omega$}  &
\multicolumn{1}{c}{$\chi^2_{\rm min}/dof$} & \multicolumn{1}{c}{$\Delta{\rm AIC}$} & \multicolumn{1}{c}{$\Delta{\rm BIC}$}\vspace{0.5mm}
\\\hline
$\Lambda$CDM  & $0.693\pm 0.006$ & $0.02265\pm 0.00022$ &$0.976\pm 0.004$& $0.293\pm 0.007$ & - & $-1$  & 39.35/38 & - & - \\
\hline
XCDM  & $0.684\pm 0.010$ & $0.02272\pm 0.00023$ &$0.977\pm 0.005$& $0.300\pm 0.009$ & - & $-0.960\pm0.033$  & 37.89/37 & -1.25 & -2.30 \\
\hline
A1  & $0.681\pm 0.011$& $0.02254\pm0.00023 $&$0.972\pm0.005$& $0.297\pm0.008$ &$0.00057\pm 0.00043 $ & $-1$  & 37.54/37 & -0.90 & -1.95 \\
\hline
A2   & $0.684\pm 0.009$& $0.02252\pm0.00024 $&$0.971\pm0.005$& $0.297\pm0.008$ &$0.00074\pm 0.00057 $ & $-1$  & 37.59/37 & -0.95 & -2.00 \\
\hline
G1 & $0.681\pm 0.011$& $0.02254\pm0.00023 $&$0.972\pm0.005$& $0.297\pm0.008$ &$0.00059\pm 0.00045 $ & $-1$  & 37.54/37 & -0.90 & -1.95 \\
\hline
G2  & $0.682\pm 0.010$& $0.02253\pm0.00024 $&$0.971\pm0.005$& $0.297\pm0.008$ &$0.00067\pm 0.00052 $ & $-1$  & 37.61/37 & -0.97 & -2.02 \\
\hline
\end{tabular}}
\end{center}
\label{tableFitPlanck}
\begin{scriptsize}
\tablecomments{Fitting results using the same data as in \cite{PlanckDE2015}.}
\end{scriptsize}
\end{table*}


We may reassess the quality fits obtained in this work from a different point of view. While the $\chi^2_{\rm min}$ value of the overall fit for any RVM and the XCDM is seen to be definitely smaller than the $\CC$CDM one, it proves very useful to reconfirm our conclusions with the help of the time-honored Akaike and Bayesian information criteria, AIC and BIC, see\,\cite{Akaike,Sugiura,Schwarz,Burnham}.
They read as follows:
\begin{equation}\label{eq:AICandBIC}
{\rm AIC}=\chi^2_{\rm min}+\frac{2nN}{N-n-1}\,,\ \ \ \ \ {\rm BIC}=\chi^2_{\rm min}+n\,\ln N\,.
\end{equation}
In both cases, $n$ is the number of independent fitting parameters and $N$ the number of data points used in the analysis.
To test the effectiveness of a dynamical DE model (versus the $\CC$CDM) for describing the overall data, we evaluate the pairwise differences $\Delta$AIC ($\Delta$BIC) with respect to the model that carries smaller value of AIC (BIC) -- in this case, the RVM's or the XCDM. The larger these differences the higher is the evidence against the
model with larger value of  AIC (BIC) -- the $\CC$CDM, in this case.
For $\Delta$AIC and/or $\Delta$BIC in the range $6-10$ one may claim ``strong evidence'' against such model; and, above 10, one speaks of ``very strong evidence''\,\citep{Akaike,Burnham}. The evidence ratio associated to rejection of the unfavored model is given by the ratio of Akaike weights, $e^{\Delta{\rm AIC}/2}$. Similarly, $e^{\Delta{\rm BIC}/2}$ estimates the so-called Bayes factor, which gives the ratio of marginal likelihoods between the two models\,\citep{AmendolaStatistics,DEBook}.

Table 1 reveals conspicuously that the $\CC$CDM appears very strongly disfavored (according to the above statistical standards) as compared to the running vacuum models. Specifically, $\Delta$AIC is in the range $17-18$ and $\Delta$BIC around $15$ for all the RVM's. These results are fully consistent and since both  $\Delta$AIC and  $\Delta$BIC are well above $10$ the verdict of the information criteria is conclusive. But there is another remarkable feature to single out at this point, namely the fact that the simple XCDM parametrization is now left behind as compared to the RVM's. While the corresponding XCDM values of $\Delta$AIC and $\Delta$BIC are also above $10$ (reconfirming the ability of the XCDM to improve the $\CC$CDM fit) they stay roughly $4$ points below the corresponding values for the RVM's. This is considered a significant difference from the point of view of the information criteria. Therefore, we conclude that the RVM's are significantly better than the XCDM in their ability to fit the data. In other words, the vacuum dynamics inherent to the RVM's seems to describe better the overall cosmological data than the effective quintessence behavior suggested by the XCDM parametrization.

Being the ratio of Akaike weights and Bayes factor  much bigger for the RVM's than for the $\CC$CDM, the former appear definitely much more successful than the latter. The current analysis undoubtedly reinforces the conclusions of our previous study\,\citep{ApJ1}, with the advantage that the determination of the vacuum parameters is here much more precise and therefore at a higher significance level. Let us stand out some of the most important differences with respect to that work: 1) To start with, we have used now a larger and fully updated set of cosmological data; 2) The selected data set is uncorrelated and has been obtained from independent analysis in the literature, see points S1-S7) above and references therein; 3) We have taken into account all the known covariance matrices among the data; 4) In this work, $h$, $\omega_b$ and $n_s$ are not fixed a priori (as we did in the previous one), we have now allowed them to vary in the fitting process. This is, of course, not only a more standard procedure, but also a most advisable one in order to obtain unbiased results. The lack of consensus on the experimental value of $h$ is the main reason why we have preferred to use an uninformative flat prior  -- in the technical sense -- for this parameter. This should be more objective in these circumstances, rather than being subjectively elicited -- once more in the technical sense -- by any of these more or less fashionable camps for $h$ that one finds in the literature, {\cite{RiessH02011}; \cite{ChenRatra2011}; \cite{Freedman2012}; \cite{WMAP9}, \cite{ACT2013}; \cite{Aubourg}; \cite{Planck2015}; \cite{RiessH0}}, whose ultimate fate is unknown at present (compare, e.g. the value from \cite{Planck2015} with the one from \cite{RiessH0}, which is $\sim 3\sigma$ larger than the former); 5) But the most salient feature perhaps, as compared to our previous study, is that we have introduced here a much more precise treatment of the CMB, in which we used not only the shift parameter, $R$, (which was the only CMB ingredient in our previous study) but the full data set indicated in S7) above, namely $R$ together with $\ell_a$ (acoustic length) and their correlations  with $(\omega_b,n_s)$.

%
\begin{figure*}
\centering
\includegraphics[angle=0,width=0.8\linewidth]{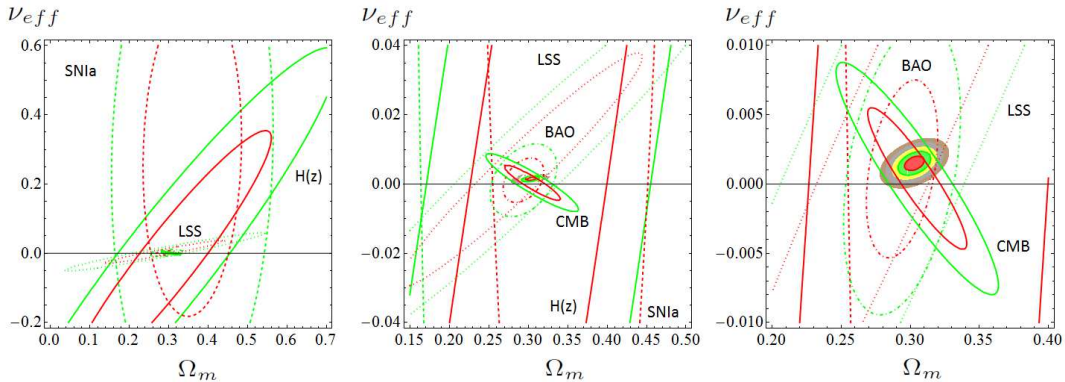}
\caption{\label{fig:A2reconstruction}%
\scriptsize Reconstruction of the contour lines for model A2, under  Planck 2015 CMB data (rightmost plot in Fig. 2) from the partial contour plots of the different SNIa+BAO+$H(z)$+LSS+BBN+CMB  data sources. The $1\sigma$ and $2\sigma$ contours are shown in all cases. For the reconstructed final contour lines we also plot the $3\sigma$, $4\sigma$ and $5\sigma$ regions.
}
\end{figure*}

Altogether, this explains the substantially improved accuracy obtained in the current fitted values of the $\nueff$ parameter as compared to \citep{ApJ1}. In particular, in what concerns points 1-3) above  we should stress that for the present analysis we are using a much more complete and restrictive BAO data set. Thus, while in our previous work we only used 6  BAO data points based on the $A(z)$ estimator (cf. Table 3 of \cite{Blake2011BAOA}), here we are using a total of 11 BAO points (none of them based on $A(z)$, see S2-S3). These include the recent results from\,\cite{GilMarin2}, which narrow down the allowed parameter space in a more efficient way, not only because the BAO data set is larger but also owing to the fact that each of the data points is individually more precise and the known correlation matrices have been taken into account. Altogether, we are able to significantly reduce the error bars with respect to the ones we had obtained in our previous work. We have actually performed a practical test to verify what would be the impact on the fitting quality of our analysis if we would remove the acoustic length $l_a$ from the CMB part of our data and replace the current BAO data points by those used in \citep{ApJ1}. Notice that the CMB part is now left essentially with the $R$-shift parameter only, which was indeed the old situation. The result is that we recover the error bars' size shown in the previous paper, which are $\sim 4-5$ times larger than the current ones, i.e. of order $\mathcal{O}(10^{-3})$. We have also checked what would be the effect on our fit if we would remove both the data on the shift parameter and on the acoustic length;  or if we would remove only the data points on LSS. The results are presented in Tables 5 and 6, respectively. We observe that the  $\Delta$AIC and $\Delta$BIC values become $2-4$ points negative. This means that the full CMB and LSS data are individually very important for the quality of the fit and that without any of them the evidence of dynamical DE would be lost. If we would restore part of the CMB effect on the fit in Table 5 by including the $R$-shift parameter in the fitting procedure we can recover, approximately, the situation of our previous analysis, but not quite since the remaining data sources used now are more powerful.

It is also interesting to explore what would have been the result of our fits if we would not have used our rather complete SNIa+BAO+$H(z)$+LSS+BBN+CMB data set and had restricted ourselves to the much more limited one used by the Planck 2015 collaboration in the paper \cite{PlanckDE2015}. {The outcome is  presented in Table 7. In contrast to \cite{Planck2015}, where no LSS (RSD) data were used, the former reference uses some BAO and LSS data, but their fit is rather limited in scope since they use only 4 BAO data points, 1 AP (Alcock-Paczynski parameter) data point, and one single LSS point, namely $f\sigma_8$ at $z=0.57$}, see details in that paper. In contradistinction to them, in our case we used 11 BAO and 13 LSS data points, some of them very recent and of high precision\,\citep{GilMarin2}.  From Table 7 it is seen that with only the data used in \cite{PlanckDE2015}  the fitting results for the RVM's are poor enough and cannot still detect clear traces of the vacuum dynamics. In particular, the $\Delta$AIC and $\Delta$BIC values in that table are moderately negative, showing that the $\Lambda$CDM does better with only these data. As stated before, not even the XCDM parametrization is able to detect any trace of dynamical DE with that limited data set, as the effective EoS is compatible with $\omega=-1$ at roughly $1\sigma$ ($\omega=-0.960\pm 0.033$). This should explain why the features that we are reporting here have been missed till now.

We complete our analysis by displaying in a graphical way the contributions from the different data sets to our final contour plots in {Figs. 1-3. We start analyzing the RVM's case.} For definiteness we concentrate on the rightmost plot for model A2 in Fig. 2, but we could do similarly for any other one {in Figs 1-2}. The result for model A2 is depicted in Fig. 5, where we can assess the detailed reconstruction of the final contours in  terms of the partial contours from the different SNIa+BAO+$H(z)$+LSS+BBN+CMB data sources. This reconstruction is presented through a series of three plots made at different magnifications. In the third plot of the sequence we can easily appraise that the BAO+LSS+CMB data subset plays a fundamental role in narrowing down the final physical region of the $(\Omega_m,\nueff)$ parameter space, in which all the remaining parameters have been marginalized over. This reconstruction also explains in very obvious visual terms why the conclusions that we are presenting here hinge to a large extent on considering the most sensitive components of the data. While CMB obviously is a high precision component in the fit, we demonstrate in our study (both numerically and graphically) that the maximum power of the fit is achieved when it is combined with the wealth of BAO and LSS data points currently available.

\begin{figure*}
\centering
\includegraphics[angle=0,width=0.52\linewidth]{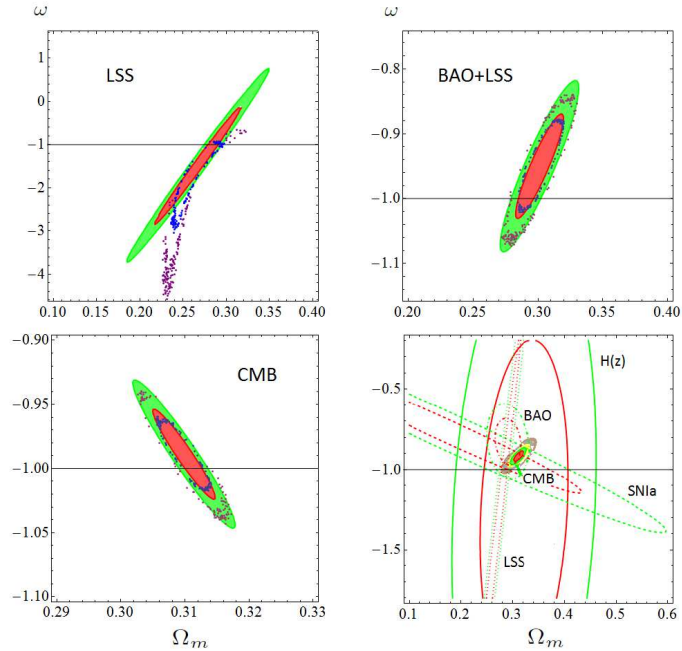}
\caption{\label{fig:XCDMreconstruction}%
{\scriptsize {\it Upper-left plot:} two-dimensional $\Omega_m-\omega$ contours at $1\sigma$ and $2\sigma$ c.l. for the XCDM, obtained with only the LSS data set. The dotted contours in blue and purple are the exact ones, whilst the red and green ellipses have been obtained using the Fisher's approximation. {\it Upper-right plot:} Same, but for the combination BAO+LSS. {\it Lower-left plot:} As in the upper plots, but for the CMB data. {\it Lower-right plot:} The Fisher's generated contours at $1\sigma$ and $2\sigma$ c.l. for all the data sets: SNIa (dotted lines), H(z) (solid lines), BAO (dot-dashed lines), LSS (dotted, very thin lines) and CMB (solid lines, tightly packed in a very small, segment-shaped, region at such scale of the plot). The exact, final, combined contours (from $1\sigma$ up to $5\sigma$) can be glimpsed in the small colored area around the center. See the text for further explanations and Fig. 3 for a detailed view.}
}
\end{figure*}

{In Fig. 6 we show the corresponding decomposition of the data contours for the XCDM model as well. In the upper-left plot we display the two-dimensional contours at $1\sigma$ and $2\sigma$ c.l. in the $(\Omega_m,\omega)$ plane, found using only the LSS data set. The elliptical shapes are obtained upon applying the Fisher matrix formalism\,\citep{AmendolaStatistics}, i.e. assuming that the two-dimensional distribution is normal (Gaussian) not only in the closer neighborhood of the best-fit values, but in all the parameter space. In order to obtain the dotted contours we have sampled the exact distribution making use of the Metropolis-Hastings Markov chain Monte Carlo algorithm \citep{Metropolis,Hastings}. We find a significant deviation from the ideal perfectly Gaussian case. In the upper-right plot we do the same for the combination BAO+LSS. The continuous and dotted contours are both elliptical, which remarkably demonstrates the Gaussian behavior of the combined BAO+LSS distribution. Needless to say the correlations among BAO and LSS data (whose covariance matrices are known) are responsible for that, i.e. they explain why the product of the non-normal distribution obtained from the LSS data and the Gaussian BAO one produces perfectly elliptical dotted contours for the exact BAO+LSS combination. Similarly, in the lower-left plot we compare the exact (dotted) and Fisher's generated (continuous) lines for the CMB data. Again, it is apparent that the distribution inferred from the CMB data in the $(\Omega_m,\omega)$ plane is a multivariate normal. Finally, in the lower-right plot we produce the contours at $1\sigma$ and $2\sigma$ c.l. for all the data sets in order to study the impact of each one of them. They have all been found using the Fisher approximation, just to sketch the basic properties of the various data sets, despite we know that the exact result deviates from this approximation and therefore their intersection is not the final answer. The final contours (up to $5\sigma$) obtained from the exact distributions can be seen in the small colored area around the center of the lower-right plot. The reason to plot it small at that scale is to give sufficient perspective to appreciate the contour lines of all the participating data. The final plot coincides, of course, with the one in Fig.3, where it can be appraised in full detail.}

{As it is clear from Fig. 6, the data on the H(z) and SNIa observables are not crucial for distilling the final dynamical DE effect, as they have a very low constraining power. This was also so for the RVM case. Once more the final contours are basically the result of the combination of the crucial triplet of BAO+LSS+CMB data (upon taking due care in this case of the deviations from normality of the LSS-inferred distribution). The main conclusion is essentially the same as for the corresponding RVM analysis of combined contours in Fig. 5, except that in the latter there are no significant deviations from the normal distribution behavior, as we have checked, and therefore all the contours in Fig. 5 can be accurately computed using the Fisher's matrix method. }

{The net outcome is that using either the XCDM or the RVM's the  signal in favor of the DE dynamics is clearly pinned down and in both cases it is the result of the combination of all the data sets used in our detailed analysis, although to a large extent it is generated from the crucial BAO+LSS+CMB combination of data sets. In the absence of any of them the signal would get weakened, but when the three data sets are taken together they have enough power to capture the signal of dynamical DE at the remarkable level of $\sim 4\sigma$.}

\section{Conclusions}

To conclude, the running vacuum models emerge as serious alternative candidates for the description of the current state of the Universe in accelerated expansion. These models have a close connection with the possible quantum effects on the effective action of QFT in curved spacetime, cf. \cite{JSPRev2013} and references therein. There were previous phenomenological studies that hinted in different degrees at the possibility that the RVM's could fit the data similarly as the $\CC$CDM, see e.g. the earlier works by \cite{BPS2009}, \cite{Grande2011}, \cite{BPS2012}, \cite{BS2014}, as well as the more recent ones by \cite{GomSolBas2015} and \cite{GomSolMNRAS}, including of course the study that precedes this work, \cite{ApJ1}.  However, to our knowledge there is no devoted work comparable in scope to the one presented here for the running vacuum models under consideration. The significantly enhanced level of dynamical DE evidence attained with them is unprecedented, to the best of our knowledge, all the more if we take into account the diversified amount of data used. Our study employed for the first time the largest updated SNIa+BAO+$H(z)$+LSS+BBN+CMB data set of cosmological observations available in the literature. Some of these data (specially the BAO+LSS+CMB part) play a crucial role in the overall fit and are substantially responsible for the main effects reported here. Furthermore, recently the BAO+LSS components have been enriched by more accurate contributions, which have helped to further enhance the signs of the vacuum dynamics. At the end of the day it has been possible to improve the significance of the dynamical hints from a confidence level of roughly $3\sigma$, as reported in our previous study \citep{ApJ1}, up to the $4.2\sigma$ achieved here. Overall, the signature of dynamical vacuum energy density seems to be rather firmly supported by the current cosmological observations. Already in terms of the generic XCDM parametrization we are able to exclude, for the first time, the absence of vacuum dynamics ($\CC$CDM) at  $4\sigma$ c.l., but such limit can be  even surpassed at the level of the RVM's and other related dynamical vacuum models, see \cite{SolCruzGomPRL} and the review \cite{JSPRev2016}.

{It may be quite appropriate to mention at this point of our analysis the very recent study of us \citep{SolGomCruzPHI}, in which we have considered the well-known Peebles \& Ratra scalar field model with an inverse power law  potential $V(\phi)\propto \phi^{-{\alpha}}$ \citep{PeeblesRatra88a,PeeblesRatra88b}, where the power ${\alpha}$ here should, of course, not be confused with a previous use of $\alpha$ for model A2 in Sect. 2). In that study we consider the response of the Peebles \& Ratra model when fitted with the same data sets as those used in the current work. Even though there are  other recent tests of that model, see e.g. the works by \cite{Samushia2009,Farooq2013,Copeland2013,Ratra2014,Samushia2014,PourtsidouTram2016}, none of them used a comparably rich data set as the one we used here. This explains why the analysis of \cite{SolGomCruzPHI} was able to  show that a non-trivial scalar field model, such as the Peebles \& Ratra model, is able to fit the observations at a level comparable to the models studied here. In fact, the central value of the ${\alpha}$ parameter of the potential is found to be nonzero at $\sim 4\sigma$ c.l., and the corresponding equation of state parameter $\omega$ deviates consistently from $-1$ also at the $4\sigma$ level. These remarkable features are only at reach when the crucial triplet of BAO+LSS+CMB data are at work in the fitting analysis of the various cosmological models. The net outcome of these investigations is that several models and parametrizations of the DE do resonate with the conclusion that there is a significant ($\sim 4\sigma$) effect sitting in the current wealth of cosmological data. The effect looks robust enough and can be unveiled using a variety of independent frameworks. }
Needless to say, compelling statistical evidence conventionally starts at $5\sigma$ c.l. and so we will have to wait for updated observations to see if such level of significance can eventually be attained. In the meanwhile the possible dynamical character of the cosmic vacuum, as suggested by the present study, is pretty high and gives hope for an eventual solution of the old cosmological constant problem, perhaps the toughest problem of fundamental physics.


\section{Acknowledgements}

We thank Gil-Mar\'in for useful discussions on BAO estimators. JS has been supported by FPA2013-46570 (MICINN), CSD2007-00042 (CPAN) and by 2014-SGR-104 (Generalitat de
Catalunya); AGV acknowledges the support of an APIF  grant of the
U. Barcelona. We are also partially supported by  MDM-2014-0369 (ICCUB).

\newcommand{\CQG}[3]{{ Class. Quant. Grav. } {\bf #1} (#2) {#3}}
\newcommand{\JCAP}[3]{{ JCAP} {\bf#1} (#2)  {#3}}
\newcommand{\APJ}[3]{{ Astrophys. J. } {\bf #1} (#2)  {#3}}
\newcommand{\AMJ}[3]{{ Astronom. J. } {\bf #1} (#2)  {#3}}
\newcommand{\APP}[3]{{ Astropart. Phys. } {\bf #1} (#2)  {#3}}
\newcommand{\AAP}[3]{{ Astron. Astrophys. } {\bf #1} (#2)  {#3}}
\newcommand{\MNRAS}[3]{{ Mon. Not. Roy. Astron. Soc.} {\bf #1} (#2)  {#3}}
\newcommand{\PR}[3]{{ Phys. Rep. } {\bf #1} (#2)  {#3}}
\newcommand{\RMP}[3]{{ Rev. Mod. Phys. } {\bf #1} (#2)  {#3}}
\newcommand{\JPA}[3]{{ J. Phys. A: Math. Theor.} {\bf #1} (#2)  {#3}}
\newcommand{\ProgS}[3]{{ Prog. Theor. Phys. Supp.} {\bf #1} (#2)  {#3}}
\newcommand{\APJS}[3]{{ Astrophys. J. Supl.} {\bf #1} (#2)  {#3}}

\newcommand{\Prog}[3]{{ Prog. Theor. Phys.} {\bf #1}  (#2) {#3}}
\newcommand{\IJMPA}[3]{{ Int. J. of Mod. Phys. A} {\bf #1}  {(#2)} {#3}}
\newcommand{\IJMPD}[3]{{ Int. J. of Mod. Phys. D} {\bf #1}  {(#2)} {#3}}
\newcommand{\GRG}[3]{{ Gen. Rel. Grav.} {\bf #1}  {(#2)} {#3}}


\end{document}